\documentclass[aps,physrev,superscriptaddress]{revtex4-2}

\usepackage{graphicx}
\usepackage{bm}
\usepackage{amsmath}
\usepackage[hidelinks]{hyperref}
\usepackage{xcolor}
\usepackage{caption}
\begin{document}


\title{\textbf{Infrared imaging of thermally-driven jets and eddies in planetary-style laboratory turbulence}}

\author{Cy S. David}
 \email{Contact author: cysdavid@ucla.edu}
 \affiliation{Department of Earth, Planetary, and Space Sciences, University of California, Los Angeles, CA 90095, USA}
 
\author{Rémy Monville}
\affiliation{Department of Earth, Planetary, and Space Sciences, University of California, Los Angeles, CA 90095, USA}

\author{Daphné Lemasquerier}
\affiliation{School of Mathematics and Statistics, University of St. Andrews, St. Andrews, United Kingdom}

\author{Lvcian Vltava}
\affiliation{Department of Earth, Planetary, and Space Sciences, University of California, Los Angeles, CA 90095, USA}

\author{Jonathan M. Aurnou}
\affiliation{Department of Earth, Planetary, and Space Sciences, University of California, Los Angeles, CA 90095,
USA}

\date{\today}

\begin{abstract}
This paper is associated with a poster winner of a 2025 American Physical Society’s Division of Fluid Dynamics
(DFD) Milton van Dyke Award for work presented at the DFD Gallery of Fluid Motion. The original poster is available online at the Gallery of Fluid Motion, \url{https://doi.org/10.1103/APS.DFD.2025.GFM.P012}
\end{abstract}


\maketitle

\section{Introduction}\label{sec:intro}
Rapidly-rotating turbulent flows in planetary fluid layers commonly feature strong axisymmetric east-west winds or ``zonal jets'' that alternate direction with latitude \citep{galperin_zonal_2019}. While the most iconic examples of these deep-seated zonal jets may be found on Saturn and Jupiter (where they are associated with bands of dark and light clouds) \citep{kaspi_jupiters_2018,cao_strong_2023,galanti_observational_2025}, these east-west flows also likely dominate the motions of icy moon subsurface oceans \citep{soderlund_ocean_2019,bire_exploring_2022,cabanes_zonostrophic_2024} and brown dwarf atmospheres \citep{zhang_atmospheric_2014,showman_atmospheric_2019,tan_atmospheric_2021}. Understanding these highly nonlinear, turbulent flows is essential to predicting the transport of heat \citep{aurnou_convective_2008,yadav_effect_2016,guervilly_multiple_2017,daniel_new_2026}, chemical species \citep{smith_tracer_2005,tan_atmospheric_2021,tan_jet_2022}, and magnetic flux \citep{aubert_steady_2005,heimpel_relationship_2011,cao_zonal_2017,wicht_dynamo_2019} across a range of astrophysical systems, with implications for planetary evolution and habitability \citep{soderlund_ocean_2019,bire_exploring_2022}.

Planetary jets in the regime of ``zonostrophic'' turbulence \citep{galperin_anisotropic_2006} are thought to arise through the anisotropization of the inverse cascade of energy in rapidly-rotating flows \citep{galperin_zonal_2019}. Vortex-stretching induced by planetary curvature (the ``topographic $\beta$-effect") supports Rossby waves, which interact with turbulent eddies and lead to an anisotropic cascade into zonal modes \citep{vallis_generation_1993}. The small-scale energy injection mechanism (that ultimately drives the jets) in the Jovian atmosphere lacks consensus \citep{kaspi_formation_2007,young_forward_2017,fletcher_how_2020,read_dynamics_2024}, with both deep convection (e.g., \citep{busse_laboratory_1976,aurnou_strong_2001,heimpel_simulation_2016}) and baroclinic instability in the shallow cloud layer (e.g., \citep{williams_jovian_2003,showman_deep_2006}) shown to be capable of driving deep-seated zonal jets. Even if a convective driving mechanism is assumed, the way that jet amplitude and spacing scales with the strength of buoyancy forcing remains challenging to predict due to (i) the difficulty in reaching extreme regimes relevant to planetary interiors \citep{gastine_scaling_2016,cheng_heuristic_2018,schwaiger_force_2019,barrois_comparison_2022,hawkins_laboratory_2023,lemasquerier_europas_2023}, (ii) the multi-stability of zonal jet configurations \citep{bouchet_rare_2019,lemasquerier_zonal_2021,simonnet_multistability_2021}, and (iii) the de-correlation of velocity fluctuations with increasing convective forcing \citep{christensen_zonal_2002,nicoski_asymptotic_2024}. Further, the mechanisms through which zonal jets alter convective heat flux and feed back on convection are debated \citep{yadav_effect_2016,guervilly_multiple_2017}. The nature of these jet--convection interactions has largely remained elusive due to the long integration times required for jets to emerge in direct numerical simulations (DNS), which must be conducted at extremely low viscosity and rapid rotation \citep{christensen_zonal_2002,heimpel_simulation_2005,heimpel_simulation_2016,yadav_effect_2016,wulff_zonal_2022,gastine_latitudinal_2023,cabanes_zonostrophic_2024,nicoski_asymptotic_2024,christensen_quenching_2024}.

Laboratory experiments have made significant progress in the study of zonal jets (see \citep{read_eddy-driven_2025,le_bars_laboratory_2026} for a review). However, most of these studies inject energy at small scales via mechanical (e.g., \citep{cabanes_laboratory_2017,lemasquerier_zonal_2021}) or electromagnetic forcing mechanisms (e.g., \citep{di_nitto_simulating_2013}) that are unrealistic for planetary atmospheres and oceans. To address this, we have designed a thermally-driven rotating laboratory experiment in which infrared (IR) imaging of the free-surface temperature field reveals the interaction of zonal jets, baroclinic instability, and turbulent convection relevant to planetary flows.

\section{Experimental setup}\label{sec:setup}
\begin{figure}
    \includegraphics[width=0.9\linewidth]{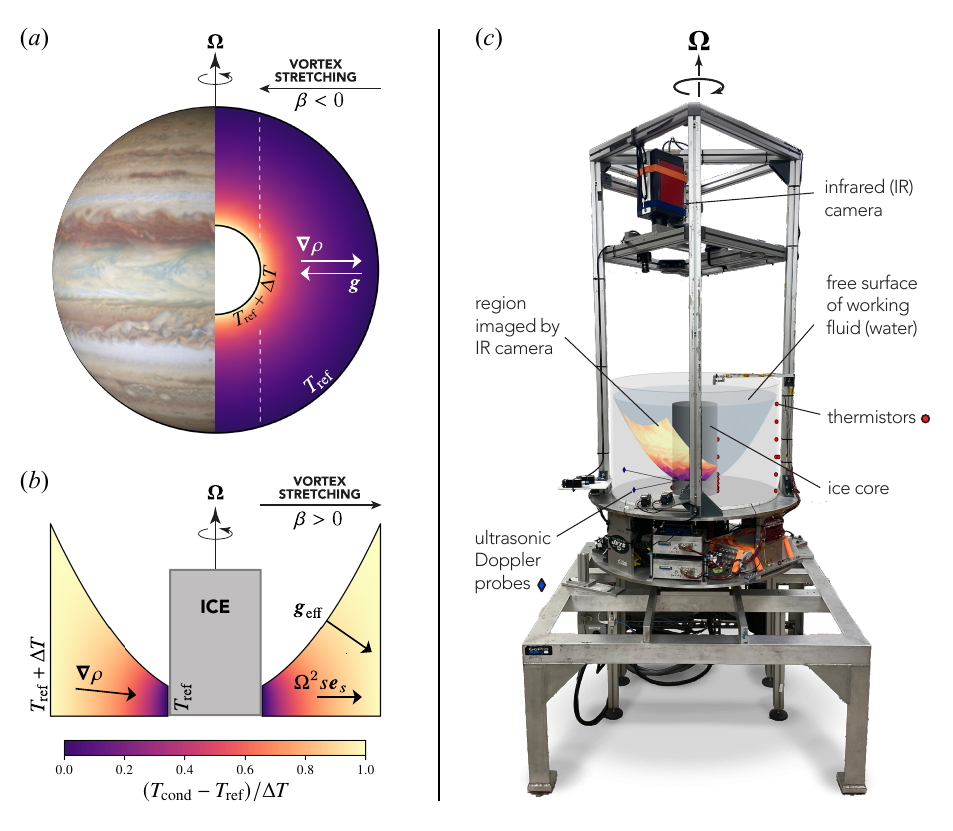}
    \caption{(a) \textit{Left half}: Image of Jupiter (credit: NASA/ESA/Goddard/UC Berkeley/JPL-Caltech/STScI) \textit{Right half}: Schematic diagram of the conductive base state in the region of a fluid planetary interior outside the tangent cylinder (dashed white line). The thickness of the layer is greatly exaggerated with respect to the depth of the jets on Jupiter. Colors correspond to the conductive temperature field for boundaries with fixed temperature difference $\Delta T$. The associated density gradient $\bm{\nabla}\rho$ is everywhere antiparallel to gravitational acceleration $\boldsymbol{g}$. (b) Analogous diagram for an open-top centrifugal convection experiment, in which the parabolic deflection of the fluid surface results in a topographic $\beta$-effect of comparable magnitude to that in planetary interiors. Ice within the central cylinder produces density gradients nearly antiparallel to the centrifugal acceleration $\Omega^2 s \bm{e}_s$. (c) Diagram of UCLA's Coreaboloid device, fitted with a 74.5 cm-diameter open-top annular channel that is rotated counterclockwise about its axis. All sensing systems are mounted in the rotating frame.}
    \label{fig:device}
\end{figure}

Figure \ref{fig:device}(a) illustrates the dynamical conditions of low-latitude flows within a planetary fluid layer (the thickness of the spherical shell is greatly exaggerated relative to the depth of Jupiter's zonal jets). Rapid rotation axializes flow structures such that the dominant form of vortex stretching occurs due to the relative change in fluid height $h$ (where $h$ is measured parallel to the rotation axis) with distance $s$ from the rotation axis (e.g., \citep{heimpel_turbulent_2007,vallis_atmospheric_2017,lemasquerier_zonal_2021}). The strength of this topographic $\beta$-effect is given by
\begin{equation}
	\beta = \frac{2\Omega}{h}\frac{\mathrm{d} h }{\mathrm{d} s},
\end{equation}
where $\Omega$ is the angular rotation rate. Colors in Figure 1(a) correspond to the conductive temperature field $T_{\text{cond}}$ for outer and inner boundaries with temperatures $T_{\text{ref}}$ and $T_{\text{ref}} + \Delta T$, respectively. Positive spherically radial density gradients $\bm{\nabla} \rho$ and negative spherically radial gravitational acceleration $\bm{g}$ result in convection, which ultimately drives zonal jets. The development of density gradients across the inner-core tangent cylinder (dashed white line) can also lead to pronounced baroclinicity ($\bm{\nabla} \rho \times \bm{g} \neq \bm{0}$) \citep{aurnou_experiments_2003,aujogue_experimental_2018}, which is unstable to nonaxisymmetric perturbations \citep{charney_dynamics_1947,eady_long_1949,charney_stability_1962}.

To simulate these conditions in the laboratory, UCLA's Coreaboloid device \citep{lonner_planetary_2022} uses an open-top annular channel rotating about its axis such that the free surface of the fluid is paraboloidally deflected under centrifugation, leading to a strong topographic $\beta$-effect as in Refs. \citep{zhang_beta-plane_2014,smith_multiple_2014,cabanes_laboratory_2017,lemasquerier_gas_2020,lemasquerier_zonal_2021,lonner_planetary_2022,lemasquerier_zonal_2023}. Figure \ref{fig:device}(b) shows the conductive temperature field within this setup for outer and inner cylindrical sidewall temperatures $T_{\text{ref}} + \Delta T$ and $T_{\text{ref}}$, respectively (with all other boundaries insulating). The density gradient now points roughly towards the rotation axis but is opposed by the centrifugal acceleration $\Omega^2 s \bm{e}_s$, leading to centrifugal convection \citep{busse_laboratory_1976,cardin_chaotic_1994,manneville_banded_1996,wang_diffusionfree_2021,lonner_planetary_2022,jiang_experimental_2022}. In addition, since the total effective gravity $\bm{g}_{\text{eff}} = -\bm{g} + \Omega^2 s \bm{e}_s$ is greatly misaligned with $\bm{\nabla} \rho$, we expect baroclinic instabilities near the inner sidewall. Both buoyancy-driven instabilities, combined with the $\beta$-effect, lead to the formation of turbulent zonal jets.

The Coreaboloid's paraboloidal configuration is shown in Figure \ref{fig:device}(c). Its annular channel, with acrylic outer sidewall of radius $R_o = 37.25$ cm and stainless steel inner sidewall of radius $R_i = R_o - L= 10.2$ cm, is filled to depth $H = 25.5$ cm with $50 ^\circ$C water (Prandtl number $\textit{Pr} =\nu/\kappa = 3.5$; throughout this work, $\nu$ and $\kappa$ denote the kinematic viscosity and thermal diffusivity of water at $50 ^\circ$C, respectively). Before rotation commences, a cylindrical core of 50 vol.\% propylene glycol--water ice is placed within the stainless steel inner cylinder and thermally coupled to the solid boundary with a thin layer of liquid water. The channel and frame are then wrapped in a fabric shroud (not shown) to minimize air drag on the water's surface, and the system is rotated at up to 80 revolutions per minute (rpm) in the counterclockwise direction (the direction of increasing azimuth, $\phi$). The free surface achieves its parabolic profile when the fluid reaches solid-body rotation, roughly at $t\approx 10 \tau_E$ \citep{odonnell_free-surface_1991}, where $\tau_E = \Omega^{-1} (2\textit{Ek})^{-1/2} H/L$ is the Ekman spin-up timescale \citep{greenspan_time-dependent_1963} and 
\begin{equation}
    \textit{Ek} = \frac{\nu}{2\Omega L^2}
\end{equation}
is the Ekman number (the ratio of viscous to Coriolis forces). We characterize the strength of the $\beta$-effect with the dimensionless average value
\begin{equation}
    \langle \tilde{\beta}\rangle_V = \frac{L}{2\Omega} \langle \beta \rangle_V, \quad \text{where} \quad \langle f \rangle_V = \frac{\int_{R_i}^{R_o} f(s) h(s) s\mathrm{d}s}{\int_{R_i}^{R_o} h(s) s\mathrm{d}s}
\end{equation}
is equal to the volume average of some field $f$ over the paraboloidal-annular domain if $f$ is axisymmetric and depth-invariant. This volume average prevents large values of $\beta(s)$ near the inner cylinder (where there is little fluid) from biasing $\langle \tilde{\beta}\rangle_V$.

Between the time when solid-body rotation is achieved ($t \approx 10 \tau_E$, typically $\sim$30 minutes) and the time when the inner ice core fully melts (up to $\sim$2.5 hours), we observe each experiment over thousands of rotation periods during which the system is in a quasi-steady state of rapidly-rotating convection (e.g., \citep{daly_convection_1980,choblet_3d_2000}), with the mean temperature gradient decreasing slowly. Thermistors within the ice core and on the inner cylinder monitor the ice temperature and heat flux across the inner sidewall. Along the inner and outer sidewalls, opposing vertical chains of 14 thermistors (red points in Figure \ref{fig:device}(c)) measure the temperature difference $\Delta T$ across the channel. These data permit instantaneous estimates of the strength of convective forcing realized in each experiment, characterized by the centrifugal Rayleigh number (e.g., \citep{jiang_experimental_2022})
\begin{equation}
    \textit{Ra}(t) =  \frac{\alpha   \Delta T(t)  \Omega ^2 R_m L^3}{ \nu  \kappa},
\end{equation}
where $\alpha$ is the thermal expansivity and $R_m = (R_i + R_o)/2$ is the mean radius of the annular gap.

Radial profiles of the radial and azimuthal flow components are estimated using ultrasonic Doppler velocimetry (UDV) probes (blue diamonds in  Figure \ref{fig:device}(c)) embedded in the outer sidewall 6.8 cm above the base of the tank. (See \citep{lonner_planetary_2022} and references therein for further details on the UDV technique and azimuthal flow estimation). We apply a low-pass filter with cutoff frequency $\omega_c = 0.2 \Omega$ to the UDV-derived azimuthal flow timeseries $u_{\phi,\text{UDV}}(s,t)$ in order to remove fast inertial modes \citep{miles_free-surface_1963,miles_free-surface_1964,scollo_resonances_2023} that are difficult to prevent in our free-surface setup (e.g., \citep{rodda_baroclinic_2018}). Then, the strength of the jets, relative to rotational effects, is quantified with the root mean square (RMS) azimuthal Rossby number defined here as
\begin{equation}\label{eqn:rossby_phi}
    \textit{Ro}_\phi(t) = \frac{1}{2\Omega L} \overline{\sqrt{\langle u_{\phi,\text{UDV}}^2\rangle_V}}, \quad \text{where} \quad \overline{f}(t) = \frac{2}{w}\int_{\max(t_0,t - w/2)}^{\min(t_1,t+w/2)} \cos^2\left[\frac{\pi(t-\tau)}{w}\right]f(\tau)\mathrm{d}\tau
\end{equation}
denotes a rolling time-mean of some timeseries $f$ on $t\in[t_0,t_1]$ using a Hann window function of width $w = 500(2\pi/\Omega)$. This smoothing operation averages over slowly propagating non-axisymmetric perturbations (i.e., Rossby waves as in \citep{lonner_planetary_2022}) and transient fluctuations in the jet structure about the long-term secular change in jet strength as the thermal forcing decreases.

Two-dimensional (2D) characterization of flow structures is achieved using high-resolution IR thermography of the free surface. Figure \ref{fig:device}(c) shows an Infratec ImageIR 8320 camera with a 640$\times$512-pixel quantum detector aimed axially downward, imaging a region covering roughly 20\% of the water's paraboloidal free surface. IR images acquired at 10 Hz reveal the radiative skin temperature (e.g., \citep{carlson_surface_2018,torres_submesoscale_2025}) over this region with $\lesssim$20~mK thermal resolution and $\lesssim$0.6~mm spatial resolution.

To assign physical coordinates to the IR images, we separately image a dot-grid calibration plate lying in the horizontal plane at $z_{\text{calib}} = 6.8$ cm above the base of the (unfilled) channel. The Discorpy package \citep{vo_discorpy_2025} is used to correct any radial lens distortion in our IR image datasets, and assign to each pixel a Cartesian coordinate ($x_{\text{calib}},y_{\text{calib}}$) in the calibration plane. Then, for each pixel, we find the point ($x,y,h(x,y)$) where the (theoretically-predicted) free surface intersects the ray between the camera lens ($x_{\text{lens}}$,$y_{\text{lens}}$,$z_{\text{lens}}$) and the point on the calibration plane ($x_{\text{calib}},y_{\text{calib}},z_{\text{calib}}$) associated with the pixel. In the following, we always plot the temperature field $T$ against ($x,y$), effecting an orthographic projection of the free-surface temperature field onto a horizontal plane (such that distances in the projected images scale uniformly with horizontal distances in physical space). Analyzing the temperature field in this 2D coordinate system is motivated by the two-dimensionalization of rapidly-rotating flows (at least on scales over which the change in surface height is sufficiently small, see \citep{calkins_three-dimensional_2013}).

\section{Thermographic observations of jet-dominated turbulence}\label{sec:observations}
We collect thermographic measurements for three experiments rotated at 40 rpm, 50 rpm, and 72 rpm ($\textit{Ek} = 8.7\times 10^{-7}, 7\times 10^{-7}, 4.9\times 10^{-7}$ and $\langle \tilde{\beta}\rangle_V = 0.5, 0.8, 1.6$, respectively). In the following, we compare IR snapshots taken at $t = 3600 (2\pi/\Omega)> 20 \tau_E$, at which time the 40, 50, and 72 rpm cases achieve values of the Rayleigh number equal to $\textit{Ra} = 2.4 \times 10^9$, $4.7 \times 10^9$, and $2.4 \times 10^{10}$, and values of the P\'eclet number (the ratio of advection to thermal diffusion) equal to $
\textit{Pe}_\phi=\textit{Ro}_\phi \textit{Ek}^{-1}\textit{Pr} = 4000$, $6300$, and $1.1 \times 10^4$, respectively. Since $\textit{Pe}_\phi \gg 1$ in all three cases, we expect IR images of the temperature field to reveal the turbulent structure of the flow.

\begin{figure}
    \includegraphics[width=0.9\linewidth]{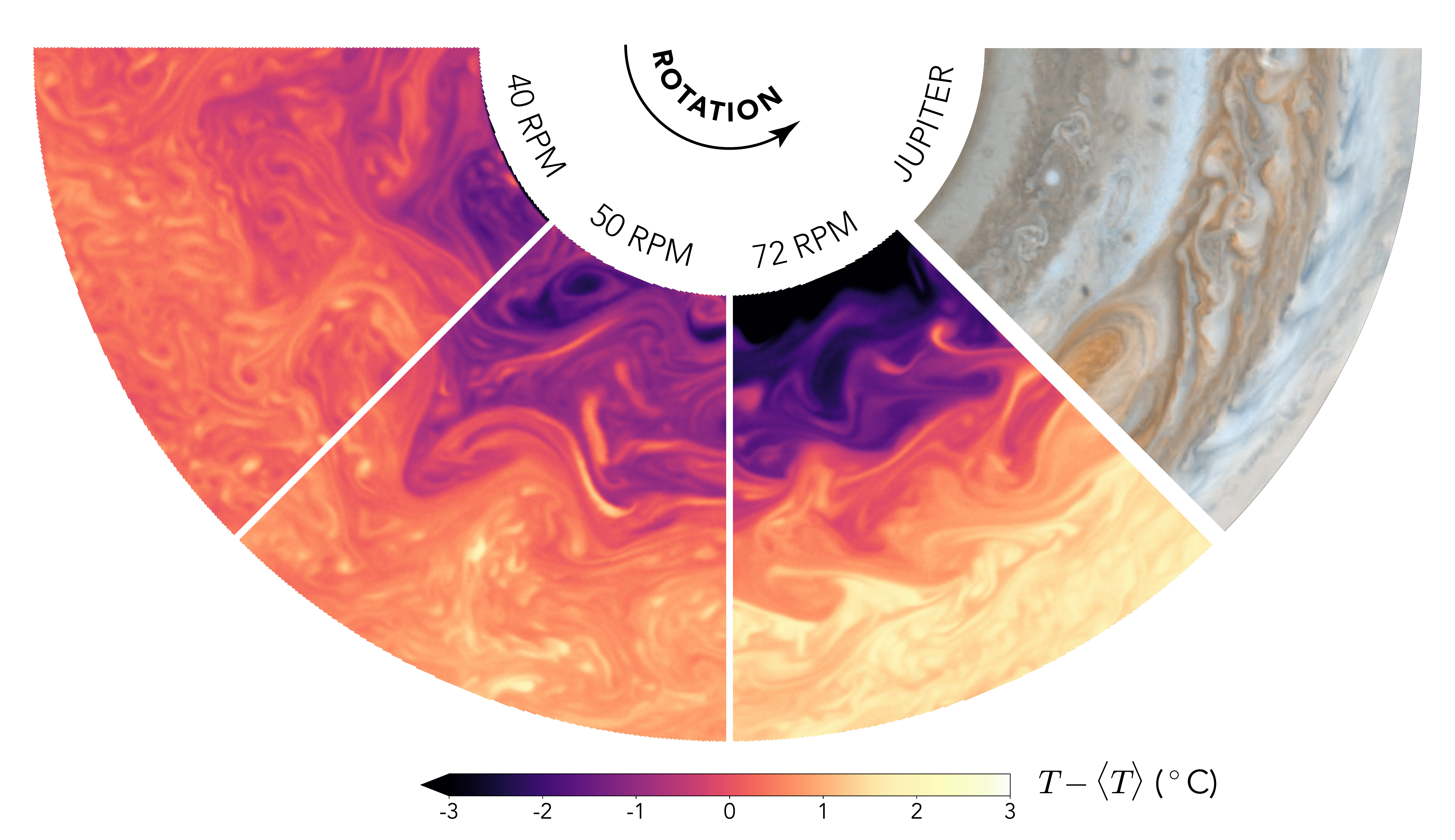}
    \caption{Top view of the infrared (IR) thermography-derived temperature field of the fluid surface (relative to an instantaneous spatial mean $\langle T \rangle$) for three laboratory experiments with increasing rotation rate. The IR images are orthographically projected from the free surface onto the horizontal plane such that distances in the images scale uniformly with horizontal distances in physical space. The inner boundary of each sectoral portion of the IR images roughly corresponds to the inner sidewall of the annular channel. Snapshots are taken at $t = 3600 (2\pi/\Omega)$. The rightmost panel shows a portion of Jupiter's southern hemisphere (credit: NASA/JPL/Space Science Institute).}
    \label{fig:sectors}
\end{figure}

Figure \ref{fig:sectors} shows sectoral portions of the surface temperature field $T$  (relative to the instantaneous spatial mean $\langle T \rangle = (\iint_{\mathcal{D}} T(s,\phi,t)s\mathrm{d}s\mathrm{d}\phi)/(\iint_{\mathcal{D}} s\mathrm{d}s\mathrm{d}\phi)$, where $\mathcal{D}$ is the locus of points $(s,\phi)$ in the horizontal plane covered by the IR images). As discussed above, the images are orthographically projected onto the horizontal plane such that the inner semicircular boundary in the figure corresponds roughly to the inner sidewall. Rotating convection maintains an interior (radial) temperature gradient (see \cite{julien_statistical_2012}) in all three cases, which becomes more pronounced with increasing rotation rate. At 40 rpm, advection of the temperature field is dominated by large eddies near the inner cylinder and strong radial stirring across roughly half of the channel width. As the rotation rate is increased to 72 rpm, $\langle\tilde{\beta}\rangle_V$ triples and strong zonal jets dominate the motion of temperature anomalies. The temperature field also develops a banded axisymmetric component resembling the zones and belts of light and dark clouds on Jupiter (rightmost panel of Figure \ref{fig:sectors}).

\begin{figure}
    \includegraphics[width=0.92\linewidth]{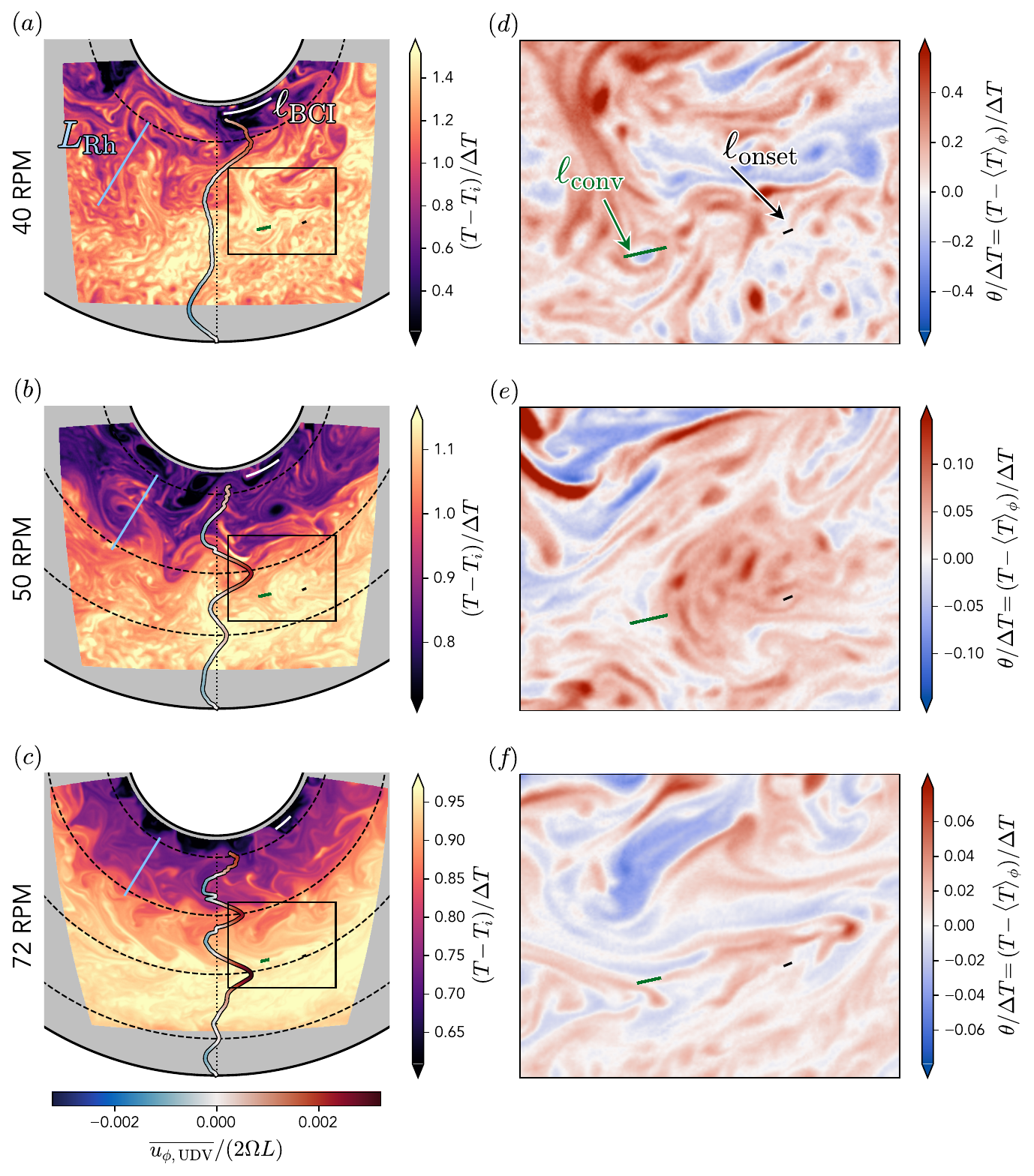}
    \caption{(a--c) Top view of the surface temperature field $T$ (relative to the thermistor-derived inner sidewall temperature $T_i$ and normalized by the sidewall temperature difference $\Delta T$) at $t = 3600 (2\pi/\Omega)$. Solid black curves indicate sidewall locations. The rolling time-average azimuthal flow profile $\overline{u_{\phi,\text{UDV}}}$ from ultrasonic Doppler velocimetry is overlaid about a vertical dotted black line (corresponding to $\overline{u_{\phi,\text{UDV}}} = 0$). Points along the profile are colored by the value of $\overline{u_{\phi,\text{UDV}}}/(2\Omega L)$ (colorbar below panel (c)). Dashed black curves correspond to the prograde peaks of $\overline{u_{\phi,\text{UDV}}}$. The predictions for the Rhines scale $L_{\text{Rh}}$ and baroclinic instability wavelength $\ell_{\text{BCI}}$ are shown with blue and white curves, respectively. (d--f) Closeup view corresponding to the black box in each panel to the left, showing the temperature deviation $\theta$ from the zonal mean $\langle T \rangle_\phi$. Predictions for the rotating convection onset scale $\ell_{\text{onset}}$ and diffusivity-free turbulent scale $\ell_{\text{conv}}$ are indicated with black and green lines, respectively.}
    \label{fig:closeups}
\end{figure}

Figures \ref{fig:closeups}(a--c) show orthographic projections of the full surface temperature field $T$ relative to the thermistor-derived inner sidewall temperature $T_i(t)$ and scaled by the sidewall temperature difference $\Delta T(t)$. Inner and outer solid black curves correspond to the channel sidewalls. The UDV-derived zonal flow profile $\overline{u_{\phi,\text{UDV}}}(s,t)$ at $t = 3600 (2\pi/\Omega)$ is overlaid about the vertical dotted line in each panel. Points along these curves are colored according to the value of $\overline{u_{\phi,\text{UDV}}}/(2\Omega L)$. The prograde ($+\phi$) peaks of $\overline{u_{\phi,\text{UDV}}}(s,t)$ are overlaid as black dashed curves.

To further illustrate the range of multi-scale physics occurring in these experiments, we consider four dynamical length-scales and compare their predictions to structures within the IR images. The largest scale is the spacing between (prograde) jets, long argued to correspond to the Rhines scale \citep{rhines_waves_1975}:
\begin{equation}\label{eqn:Rhines}
    L_{\text{Rh}} = 2\pi\sqrt{\frac{2\textit{Ro}_\phi}{\langle \tilde{\beta}\rangle_V}} L,
\end{equation}
which can be interpreted as the scale at which the advection of vorticity is balanced by the $\beta$-effect (vortex stretching) \citep{vallis_atmospheric_2017}. The RMS total velocity is typically used to define the Rhines scale \citep{rhines_waves_1975,scott_structure_2012}. However, the RMS zonal flow is roughly 2--4 times larger than the RMS radial flow in these experiments, so we use $\textit{Ro}_\phi$ in Equation (\ref{eqn:Rhines}) for simplicity. In Figures \ref{fig:closeups}(a--c), the Rhines scale $L_{\text{Rh}}$ is overlaid as light blue line segments. At 40 rpm, the jets are weak and incoherent, and $L_{\text{Rh}}$ roughly matches the wavelength of the undulations in the rolling time-average zonal flow profile $\overline{u_{\phi,\text{UDV}}}$. At 50 and 72 rpm, the jets are stronger and more coherent, and there is excellent agreement between $L_{\text{Rh}}$ (light blue lines), the spacing of prograde jets (dashed black curves), and the banding of the temperature field.

The next largest scale is that of the cold eddies near the inner cylinder, associated with baroclinic instability induced by the inclination of density gradients with respect to the effective gravity in rotating flows \citep{charney_dynamics_1947,eady_long_1949,charney_stability_1962,molemaker_baroclinic_2005}. The most unstable linear mode has a wavelength $\ell_{\text{BCI}}$ equal to 3.9 times the Rossby radius of deformation \citep{eady_long_1949,smyth_instability_2019,moscoso_low-cost_2023}, which we estimate using the thermistor-derived vertical temperature gradient $\partial_z T(s=R_i)$ at the inner sidewall:
\begin{equation}
    \ell_{\text{BCI}} = 3.9 \frac{h(s=R_i)}{2\Omega}\sqrt{\alpha g \partial_z T(s=R_i)}.
\end{equation}
In Figure \ref{fig:closeups}(a--c), the predicted scale $\ell_{\text{BCI}}$ (white curves) matches the size of cold eddies at the inner cylinder (nonaxisymmetric black features in the temperature field), and shrinks as the rotation rate increases.

Whereas baroclinic instability involves the exchange of light and dense fluid in the direction normal to the local effective gravity (e.g., in the $+s$ direction near the inner cylinder, where $\bm{g}_{\text{eff}}$ is close to vertical), convection involves the motion of dense fluid parallel to the effective gravity (e.g., in the $+s$ direction in the region nearer to the outer cylinder, where $\bm{g}_{\text{eff}}$ is close to horizontal) \citep{callies_baroclinic_2018,kang_modulation_2023}. These differences lead to a change in dynamics (and dominant length-scales) with radius: structures in the surface temperature field associated with centrifugal convection are expected closer to the outer cylinder than those associated with baroclinic instability. A further difference between convection and baroclinic instability involves the importance of viscosity. While diffusive effects are absent from the dominant Coriolis-buoyancy balance that sets $\ell_{\text{BCI}}$, viscous forces are required to overcome the constraining influence of the $\beta$-effect on convection \citep{busse_asymptotic_1986,calkins_three-dimensional_2013}, and thus the convective onset scale necessarily involves $\nu$.

The onset of rotating convection is associated with nonaxisymmetric temperature perturbations whose wavelength is proportional to
\begin{equation}
    \ell_{\text{onset}} = \textit{Ek}^{1/3}L
\end{equation}
\citep{chandrasekhar_hydrodynamic_1961,busse_asymptotic_1986,calkins_three-dimensional_2013}. Panels (d--f) of Figure \ref{fig:closeups} show temperature deviations $\theta$ about the zonal mean $\langle T \rangle_\phi = (2\pi)^{-1}\int_0^{2\pi} T(s,\phi,t)\mathrm{d}\phi$ in a small region (away from the inner cylinder) indicated by the black box in each of panels (a--c), respectively. In Figure \ref{fig:closeups}(d), $\ell_{\text{onset}}$ matches the smallest scale of the temperature anomalies. These convective perturbations grow and are thought to nonlinearly saturate through a balance of inertial, Coriolis, and buoyancy forces, which predicts a diffusivity-free convective Rhines scale (see \citep{aubert_systematic_2001}):
\begin{equation}
    \ell_{\text{conv}} = \langle\tilde{\beta}\rangle_V^{-1}(\textit{Ra}\textit{Ek}^2/\textit{Pr})^{1/2} L,
\end{equation}
shown with green lines in Figure \ref{fig:closeups}(d--f). Though the estimates for $\ell_{\text{onset}}$ and $\ell_{\text{conv}}$ adequately identify the structures in panels (d--f) as convective, the quantitative relationship between the control parameters and the characteristic temperature anomaly length-scale in this system requires further investigation. Definitive scaling relations for length-scales in geophysically-relevant regimes of rotating convection remain elusive \citep{guervilly_turbulent_2019,oliver_small_2023} and are essential to predictions of heat transport and magnetic induction in the fluid layers of stars, planets, and moons.

In this work, we show that the key hydrodynamic phenomena expected in rapidly-rotating planetary fluid layers --- convection, baroclinic instability, and jets \citep{vallis_atmospheric_2017,finlay_gyres_2023,soderlund_physical_2024,read_dynamics_2024} --- coexist in free-surface laboratory experiments, as revealed by IR thermographic sensing of the surface temperature $T$. The axisymmetric component of $T$ exhibits strong radial gradients at prograde jet peaks, while the non-axisymmetric temperature anomalies $\theta$ are tied to the baroclinic and convective eddies that ultimately drive these zonal flows.

\begin{acknowledgments}
C.S.D. is supported by the NSF GRFP
via award DGE-2444110. R.M. and J.M.A. acknowledge support from NASA PSIE (Precursor Science Investigations for Europa, grant \#80NSSC24K0400). D.L. acknowledges support from UK Research and Innovation via a Future Leaders Fellowship (grant MR/Y01605X/1).
\end{acknowledgments}

\bibliography{references}

\end{document}